\documentclass[fleqn,usenatbib,useAMS]{mnras}
\usepackage[T1]{fontenc}
\usepackage[utf8]{inputenc}

\DeclareRobustCommand{\VAN}[3]{#2}
\let\VANthebibliography\thebibliography
\def\thebibliography{\DeclareRobustCommand{\VAN}[3]{##3}\VANthebibliography}

\usepackage{anyfontsize}
\usepackage{acro}
\usepackage{graphicx}
\usepackage{tabularx}
\usepackage{multicol}
\usepackage{bm}
\usepackage{pdflscape}
\usepackage{amsmath,amsfonts}
\usepackage[normalem]{ulem}
\usepackage{color}
\usepackage{xcolor}
\usepackage{natbib}
\usepackage{newtxtext,newtxmath}
\usepackage{threeparttable}
\usepackage[capitalize,noabbrev]{cleveref}
\usepackage{orcidlink}
\usepackage{xparse}

\DeclareOldFontCommand{\rm}{\normalfont\rmfamily}{\mathrm}
\DeclareOldFontCommand{\sf}{\normalfont\sffamily}{\mathsf}
\DeclareOldFontCommand{\tt}{\normalfont\ttfamily}{\mathtt}
\DeclareOldFontCommand{\bf}{\normalfont\bfseries}{\mathbf}
\DeclareOldFontCommand{\it}{\normalfont\itshape}{\mathit}
\DeclareOldFontCommand{\sl}{\normalfont\slshape}{\@nomath\sl}
\DeclareOldFontCommand{\sc}{\normalfont\scshape}{\@nomath\sc}
\DeclareRobustCommand*\cal{\@fontswitch\relax\mathcal}
\DeclareRobustCommand*\mit{\@fontswitch\relax\mathnormal}

\newcommand{\units}[1]{\ensuremath{\text{\,#1}}}

\newcommand{\Msun}{\ensuremath{{M}_{\sun}}}

\newcommand{\Zsun}{\ensuremath{{Z}_{\sun}}}

\newcommand{\Mzams}{\ensuremath{{M}_\text{ZAMS}}}
\newcommand{\compactness}[1]{\ensuremath{\xi_{#1}}}

\newcommand{\punct}[1]{\ensuremath{\text{\,#1}}}

\DeclareAcronym{EOS}{short=EoS,long=equation of state,short-plural-form=EoS,long-plural-form=equations of state}
\DeclareAcronym{NSE}{short=NSE,long=nuclear statistical equilibrium}
\DeclareAcronym{QSE}{short=QSE,long=quasi-equilibrium}

\DeclareAcronym{CCSN}{short=CCSN,long=core-collapse supernova,short-plural-form=CCSNe,long-plural-form=core-collapse supernovae}
\DeclareAcronym{MR-SN}{short=MR-SN,long=magneto-rotational supernova,short-plural-form=MR-SNe,long-plural-form=magneto-rotational supernovae}
\DeclareAcronym{SESN}{short=SESN,long=stripped-envelope supernova,short-plural-form=SESNe,long-plural-form=stripped-envelope supernovae}
\DeclareAcronym{USSN}{short=USSN,long=ultra-stripped supernova,short-plural-form=USSNe,long-plural-form=ultra-stripped supernovae}
\DeclareAcronym{SN}{short=SN,long=supernova,short-plural-form=SNe,long-plural-form=supernovae}
\DeclareAcronym{SNR}{short=SNR,long=supernova remnant,short-plural-form=SNRs,long-plural-form=supernova remnants}

\DeclareAcronym{PNS}{short=PNS,long=proto-neutron star,long-plural-form=proto-neutron stars}
\DeclareAcronym{NS}{short=NS,long=neutron star,long-plural-form=neutron stars}
\DeclareAcronym{BH}{short=BH,long=black hole,long-plural-form=black holes}

\DeclareAcronym{ZAMS}{short=ZAMS,long=zero-age main sequence}
\DeclareAcronym{RSG}{short=RSG,long=red supergiant,long-plural-form=red supergiants}
\DeclareAcronym{BSG}{short=BSG,long=blue supergiant,long-plural-form=blue supergiants}
\DeclareAcronym{HeS}{short=HeS,long=helium star,long-plural-form=helium stars}
\DeclareAcronym{AGB}{short=AGB,long=asymptotic giant branch}

\DeclareAcronym{SASI}{short=SASI,long=standing accretion shock instability}
\DeclareAcronym{RLOF}{short=RLOF,long=Roche lobe overflow}
\DeclareAcronym{GCE}{short=GCE,long=galactic chemical evolution}
\DeclareAcronym{IMF}{short=IMF,long=initial mass function}
\DeclareAcronym{ISM}{short=ISM,long=interstellar medium}

\DeclareAcronym{Si/O}{short=Si/O,long=silicon-oxygen}
\DeclareAcronym{CO/He}{short=C+O/He,long=carbon-oxygen-helium}
\DeclareAcronym{H/He}{short=H/He,long=hydrogen-helium}

\newcommand{\bRSG}{12.6b}
\newcommand{\bHeS}{13.8b}
\newcommand{\sRSG}{12.6s}
\newcommand{\sHeS}{13.8s}

\definecolor{cyan}{rgb}{0,0.9,0.9}
\definecolor{orange}{rgb}{0.9,0.5,0}
\definecolor{magenta}{rgb}{1,0,1}
\definecolor{purple}{rgb}{0.8,0.4,0.8}
\definecolor{darkgreen}{rgb}{0.0,0.5,0.0}

\newcommand{\code}[1]{\textsc{#1}}
\newcommand{\flash}{\code{FLASH}}
\newcommand{\grid}{\code{GR1D}}
\newcommand{\stir}{\code{STIR}}
\newcommand{\nulib}{\code{NuLib}}

\newcommand{\mesa}{\code{MESA}}
\newcommand{\winnet}{\code{WinNet}}
\newcommand{\reaclib}{\code{JINA Reaclib}}
\newcommand{\chimera}{\code{CHIMERA}}

\NewDocumentCommand{\nucleus}{m m o}{%
  \IfNoValueTF{#3}
    {#2}%
    {\ensuremath{^{#3}}#2}%
}

\NewDocumentCommand{\hydrogen}{o}{\nucleus{hydrogen}{H}[#1]}
\NewDocumentCommand{\helium}{o}{\nucleus{helium}{He}[#1]}
\NewDocumentCommand{\carbon}{o}{\nucleus{carbon}{C}[#1]}
\NewDocumentCommand{\oxygen}{o}{\nucleus{oxygen}{O}[#1]}
\NewDocumentCommand{\neon}{o}{\nucleus{neon}{Ne}[#1]}
\NewDocumentCommand{\magnesium}{o}{\nucleus{magnesium}{Mg}[#1]}
\NewDocumentCommand{\silicon}{o}{\nucleus{silicon}{Si}[#1]}
\NewDocumentCommand{\sulphur}{o}{\nucleus{sulphur}{S}[#1]}
\NewDocumentCommand{\argon}{o}{\nucleus{argon}{Ar}[#1]}
\NewDocumentCommand{\calcium}{o}{\nucleus{calcium}{Ca}[#1]}
\NewDocumentCommand{\titanium}{o}{\nucleus{titanium}{Ti}[#1]}
\NewDocumentCommand{\chromium}{o}{\nucleus{chromium}{Cr}[#1]}
\NewDocumentCommand{\iron}{o}{\nucleus{iron}{Fe}[#1]}
\NewDocumentCommand{\nickel}{o}{\nucleus{nickel}{Ni}[#1]}
\NewDocumentCommand{\zinc}{o}{\nucleus{zinc}{Zn}[#1]}

\title[
    Nucleosynthesis from supernovae in binaries
]{
    It Takes Two: nucleosynthesis in core-collapse supernova simulations of binary-star progenitors
}

\author[G. Jalin et al.]{
Gaétan Jalin$^{\orcidlink{0009-0002-9632-4706}}$,$^{1}$\thanks{E-mail: gaetan.jalin@tu-darmstadt.de}
Almudena Arcones$^{\orcidlink{0000-0002-6995-3032}}$,$^{1,2,3}$
Evan P. O'Connor$^{\orcidlink{0000-0002-8228-796X}}$,$^{4}$
Andrea Ercolino,$^{5}$
\newauthor
\ and Norbert Langer$^{5,6}$
\\
$^{1}$Institut für Kernphysik, Technische Universität Darmstadt, Schlossgartenstr. 2, Darmstadt 64289, Germany\\
$^{2}$GSI Helmholtzzentrum für Schwerionenforschung GmbH, Planckstr. 1, Darmstadt 64291, Germany\\
$^{3}$Max-Planck-Institut für Kernphysik, Saupfercheckweg 1, Heidelberg 69117, Germany\\
$^{4}$The Oskar Klein Centre, Department of Astronomy, Stockholm University, SE-106 91 Stockholm, Sweden \\
$^{5}$Argelander Institut für Astronomie, Auf dem Hügel 71, DE-53121 Bonn, Germany \\
$^{6}$Max-Planck-Institut für Radioastronomie, Auf dem Hügel 69, DE-53121 Bonn, Germany
}

\date{Accepted XXX. Received YYY; in original form ZZZ}

\pubyear{\the\year{}}

\begin{document}
\label{firstpage}
\pagerange{\pageref{firstpage}--\pageref{lastpage}}
 \maketitle

\begin{abstract}
Binary interaction is common amongst massive stars and can significantly alter their pre-supernova structure and composition.
This has important consequences for the nucleosynthetic yields and observable properties of the resulting supernovae.
However, binary-evolved progenitors remain largely unexplored in modern neutrino-driven core-collapse supernova simulations.
In this work, we present one- and two-dimensional core-collapse supernova simulations of single- and binary-star progenitors.
We investigate the impact of binary interaction on the explosion and examine the effect of in-situ networks on the dynamics and resulting nucleosynthesis relative to post-processing approaches.
Our progenitors highlight the impact of binary interaction on the pre-supernova core structure of the star and its significance for explodability.
We find that binary progenitors are predicted to be generally more explodable than single stars with the same initial mass, indicating that the supernova outcome cannot be well correlated with the initial mass.
However, a systematic comparison of the explosion outcomes of binary- and single-star progenitors from self-consistent simulations is still required to confirm this trend.
Energy generation from nuclear reactions, as well as the abundances of key isotopes such as \nickel[56], can be well approximated by a small $\alpha$-chain network.
However, there are notable discrepancies in the production of \titanium{}, \chromium{}, and \iron{} isotopes which remain sensitive to the network size and thermodynamic history.
\end{abstract}
\acresetall



\section{Introduction} \label{sec:intro}

\Ac{CCSN} explosions are essential contributors to the chemical enrichment of the Universe, releasing into the \acl{ISM} the products of stellar fusion and explosive nucleosynthesis.
They mark the final stage of the evolution of massive stars and leave behind the \ac{NS} remnants that, if in binaries, may later merge and further enrich the Universe in heavy elements through the rapid neutron-capture process (\emph{r}-process).

Since the seminal work of \citet{ColgateWhite1966}, considerable progress has been achieved in the numerical modelling of \acp{CCSN}.
Rapid advances in computational resources and numerical methods over the last few decades have allowed the production of increasingly sophisticated multidimensional simulations of neutrino-driven \acp{CCSN} \citep[e.g.][]{Tamborra2014,Lentz2015,Janka2016,OConnor2018b,Burrows2020,Shibagaki2024,Vartanyan2025}, including accurate neutrino transport schemes \citep[see][for a detailed review]{Mezzacappa2020}, general relativity \citep[e.g.][]{Kuroda2012,OConnor2018a}, high-density \acp{EOS} \citep[e.g.][]{Hempel2012,Schneider2019,Yasin2020}, and magnetic fields \citep[e.g.][]{Mosta2015,Obergaulinger2018,Kuroda2020,Bugli2021,Kovalenko2026}.
Yet, the puzzle is far from complete, and many challenges and sources of uncertainty remain.

In particular, \ac{CCSN} simulations fundamentally depend on the stellar progenitors used to initiate them, which are typically calculated assuming isolated single stars.
However, it is now well established that the majority of massive stars interact with a companion during their evolution \citep{Sana2012,Moe2017,Sana2025} and that this interaction can dramatically alter the resulting progenitor compared to single-star evolution \citep{Langer2012,Laplace2021}.
For example, a large fraction of massive binaries undergo stable mass transfer through \ac{RLOF}, which results in the primary star being partially or completely stripped of its \hydrogen{} envelope, while the secondary accretes part of the \hydrogen{}-rich material.
As the stars evolve to core-collapse, the explosion of the primary may be observed as a Type Ib, Ic, or IIb \ac{SN}, also referred to as \ac{SESN}, while the secondary may instead produce a \hydrogen{}-rich explosion \citep{Renzo2020,Ercolino2026}.
Furthermore, an initially low-mass secondary star, that would be too small to undergo core-collapse, can accrete enough mass from the primary to form an iron core and produce an explosion \citep{Wellstein2001,Zapartas2021}.
Other binary evolution pathways such as mergers and phases of unstable mass transfer exist and may result in even more exotic \ac{SN} progenitors.
It is therefore essential to include binary-evolved progenitors in \ac{CCSN} simulations.

While \citet{Brown2001} showed that binary mass transfer affects the explosion properties of the mass donors, the systematic exploration of corresponding supernova explosions models has begun only recently.
\citet{Tauris2015} produced a number of \ac{SESN} progenitors evolved in a tight binary with a \ac{NS} companion, which are of particular interest in the context of the formation of double neutron star systems and gravitational wave detection.
Some of these models were later exploded by \citet{Muller2018,Muller2019}; however, the progenitors had not been evolved self-consistently.
Instead, they assumed prior envelope stripping from a common envelope phase, thus starting the evolution with a \ac{HeS} and a point mass \ac{NS} companion.
The models exploded by \citet{Suwa2015} and \citet{Ertl2020} were produced using a similar setup; these studies focused on bare \helium[] and \carbon[]-\oxygen[] cores as progenitors of \acp{SESN}.
The full variety of binary evolution is now being explored in greater detail by, e.g. \citet{Laplace2021,Farmer2023,Ercolino2024,Gilkis2025,Jin2026}, over a wide range of initial orbital separations, mass ratios, initial masses, and metallicities.
However, self-consistent simulations of the core-collapse of binary-evolved stars are still scarce \citep[but see][]{Vartanyan2021,Nakamura2022}.

Another important limitation of current \ac{CCSN} simulations is their limited ability to predict observable properties of \acp{SN} and \acp{SNR}.
This requires a detailed account of the nucleosynthesis and its spatial distribution, as well as long-term multidimensional simulations that follow the ejected material to shock breakout.
The radioactive isotopes \titanium[44] and \nickel[56] are strong examples, and accurately capturing their evolution in \ac{CCSN} simulations is key to producing theoretical observables that can be used to constrain the properties of the central engine and its connection to the pre-\ac{SN} progenitor.
Furthermore, \citet{Muller2018} highlighted that Rayleigh-Taylor mixing of radioactive isotopes in the stellar envelope during shock expansion may have an impact on the spectroscopic signature of \acp{SESN}.
In general, long-term simulations to shock breakout reveal the importance of \nickel[56] in shaping the shock morphology \citep{Vartanyan2025}.
As the shock crosses the \ac{H/He} interface, the \iron[]-rich ejecta mixes with the stellar envelope, forming fast-moving \nickel[56] bullets and leading to significant variations in the \ac{SN} light curve \citep{Vartanyan2026}.
To address these effects, it is necessary to include an appropriate nuclear network in \ac{CCSN} simulations.

The problem of the treatment of nuclear reactions in simulations of \acp{CCSN} is a long-standing one.
Evolving large nuclear networks together with the hydrodynamics in multidimensional \ac{CCSN} simulations becomes extremely computationally expensive already for networks 
which include more than $100$--$200$ species.
For this reason, most simulations neglect nuclear reactions altogether and keep the existing composition from the progenitor or assume \ac{NSE}.
Instead, the nuclear reactions can be calculated in post-processing using tracer particles and an external nuclear network.
However, this approach does not account for the feedback from nuclear energy generation on the hydrodynamics, which may alter both the explosion and the nucleosynthesis \citep[see, e.g.][]{Harris2017,Navo2023}.

A more attractive, computationally affordable alternative is to use a reduced $\alpha$-chain network to track only $10$--$20$ species.
Such networks have long been used to estimate the impact of nuclear burning on the explosion dynamics \citep[see, e.g.][]{Muller1986,Timmes2000a,Kifonidis2003,Nakamura2014,Bruenn2016,Wongwathanarat2017,Bollig2021,Nakamura2022}.
They are usually considered good approximations because they are nearly complete for \carbon[] and \oxygen[] burning, and because explosive nucleosynthesis in conditions close to symmetric matter ($Y_e \simeq 0.5$), common in \ac{SN} explosions, tends to keep the nuclear flow close to the $\alpha$ chain.
However, they are too small to capture nuclear reactions under more neutron- or proton-rich conditions and cannot generally provide detailed nucleosynthesis yields.

A few simulations have recently been produced using larger networks.
\citet{Harris2017} and \citet{Sandoval2021} produced a handful of 2D simulations as well as one 3D simulation using the 150- and 160-species networks implemented in the \chimera{} code. \citet{Navo2023} performed 2D simulations using a reduced, effectively 242-species network, which accounts for the rates of 242 species while explicitly evolving only 94 thanks to the assumption of steady-state nuclei.
Nonetheless, even such comparatively large networks are in principle insufficient to capture all the nuclear reactions contributing to the energy generation and nucleosynthesis.
The missing reactions lead to a stagnation of the nuclear flow at the edges of the network, and artificially increase the abundances of the bottleneck species \citep{Sandoval2021,Farmer2023,Navo2023}.

The ability to simulate the explosions of binary-evolved stars together with a nuclear network is particularly important to connect \ac{CCSN} theory to observations.
Furthermore, the resulting nucleosynthesis predictions can directly impact other fields such as \ac{GCE} modelling \citep{Prantzos2018,Kobayashi2020,Jost2024}.
Current state-of-the-art \ac{GCE} models typically neglect the impact of binarity due to the lack of explosive nucleosynthesis data from self-consistent \ac{CCSN} models of binary-evolved stars.
It is therefore necessary to perform \ac{CCSN} simulations of large grids of binary progenitors and derive the explosive yields.

In this work, we investigate the explosion and nucleosynthesis of the primary stars of binary systems by performing multidimensional \ac{CCSN} simulations.
We compare the explosion properties with those of single-star progenitors with the same initial masses to assess the impact of binary evolution on the final outcome.
The nucleosynthesis is calculated in the simulations by employing the same networks as in \citet{Navo2023}, including feedback from nuclear heating, and compared to post-processing calculations using a large external network.

This paper is organised as follows: In \cref{sec:methods}, we introduce the numerical method, input physics, and progenitor stars used to simulate our \ac{CCSN} models; we discuss and compare the differences between binary- and single-star progenitors based on their pre-\ac{SN} core structure in \cref{sec:explodability}. In \cref{sec:results}, we present the results of our simulations by first comparing single-star models to binary-evolved ones, and then the nucleosynthesis results from the binary progenitors. Finally, we present our conclusions in \cref{sec:outro}.

\section{Methods} \label{sec:methods}

\subsection{Simulation setup} \label{sec:methods_hydro}

We use a modified version of the \flash{} code \citep[version 4;][] {Fryxell2000,Dubey2009} specialised for simulations of \acp{CCSN} \citep{Couch2013,Couch2014,OConnor2018a,OConnor2018b}.
This version solves Newtonian hydrodynamics with a general relativistic correction to the gravitational potential \citep[][Case~A]{Marek2006} and uses an energy-dependent two-moment neutrino transport with an M1 analytic closure \citep{OConnor2018a}.
The neutrino transport uses 12 logarithmically spaced energy groups and evolves three neutrino species: electron neutrinos ($\nu_e$), electron antineutrinos ($\bar{\nu_e}$), and a third species collecting all heavy-lepton neutrinos and antineutrinos ($\nu_x$).
Neutrino opacities are generated with the neutrino opacity library \nulib{} \citep{OConnor2015} following the baseline set from \citet{OConnor2018b}.
We also include virial and effective many body corrections \citep{Horowitz2017}, and inelastic neutrino--electron scattering following \citet{Bruenn1985}.

Our simulations are performed using adaptive mesh refinement;
the 1D models use spherical geometry assuming spherical symmetry,
and the 2D models use cylindrical geometry assuming axisymmetry around the polar axis.
This allows us to maintain a higher resolution inside the \ac{PNS} and close to shocks while avoiding severe computational bottlenecks.
We achieve a finest grid spacing of $\sim305\units{m}$ for the 1D simulations, and $\sim355\units{m}$ for the 2D simulations.

Multidimensional effects, such as convection and the \ac{SASI}, are known to enhance neutrino heating in the gain region and can be a deciding factor between a failed and a successful explosion \citep{Marek2009,Murphy2013,Couch2015,Radice2016}.
One-dimensional simulations intrinsically lack these effects and generally fail to produce explosions self-consistently \citep{Rampp2000,Liebendorfer2001}, except for a few low-mass progenitors \citep{Kitaura2006}.
To palliate this problem, it is customary to introduce some parametrisation to artificially trigger the explosion.
For example, the `PUSH' \citep{Perego2015} or the \stir{} methods \citep{Couch2020} are common approaches.
In this work, we employ a heating factor: a simple multiplier to the energy deposited by neutrino-matter interactions in the gain layer.
We use a heating factor of $2.8$ for all our 1D models. This value ensures that all our progenitors explode while producing results that compare well with both observations and multidimensional simulations \citep{Jost2024}.

\subsection{Nuclear network and equation of state} \label{sec:methods_network_eos}

We have implemented the two reduced nuclear reaction networks presented in \citet{Navo2023} in \flash{}.
The smaller network (hereafter `RN16') is a 16-species $\alpha$-chain nuclear reaction network which includes neutrons, protons, \helium[4], \iron[54], and all $\alpha$-nuclei from \carbon[12] to \nickel[56].
The extra \iron[54] nucleus allows the network to converge in slightly neutron-rich environments.
The larger network (hereafter `RN94') is composed of 94 species from free neutrons and protons to \nucleus{molybdenum}{Mo}[92].
These networks assume additional species to be in steady-state.
These species are not explicitly evolved by the network but contribute to the evolution of the main species by connecting intermediate reactions in which they are involved.
With this treatment, the RN16 and RN94 networks effectively correspond to a 25- and 242-nuclei reaction networks, respectively, but are much less computationally expensive (see \citealt{Navo2023} for further details).

The implementation of the nuclear network in our \ac{CCSN} simulations is tied to the \ac{EOS}, whose formulation depends on the local thermodynamic conditions.
\flash{} employs a hybrid \ac{EOS} \citep{Witt2021} in which the so-called Helmholtz \ac{EOS} \citep{Timmes1999a,Timmes2000b} is used at low densities and transitions to a tabulated \ac{EOS} for nuclear matter at higher densities.

In the Helmholtz \ac{EOS} regime the network operates normally.
The composition is initially given by the progenitor model, and is burned hydrostatically at the end of each hydrodynamic time step.
The updated composition can be used by the Helmholtz \ac{EOS} for the next step and, if nuclear heating is enabled, the energy released or absorbed by nuclear reactions is added to the internal energy of the system.

In the high-density \ac{EOS} regime, the composition and thermodynamic quantities are precomputed with the \ac{EOS} table; therefore, the internal network is not evolved.
Nevertheless, the composition must remain consistent across the interface between the two \ac{EOS} to ensure a smooth transition when material enters or leaves the high-density regime.
Prior to the implementation of the network, the composition was obtained directly from the \ac{EOS} table and frozen upon flowing out of the high-density regime into the Helmholtz regime.
With the inclusion of the network, the composition of the network must be initialised before entering the Helmholtz regime.
We thus define a region $T_\text{high} > T \geq T_\text{NSE}$ within which the composition of the network is calculated assuming \ac{NSE}.
This ensures that all cells in \ac{NSE} have their composition correctly initialised when cooling below $T_\text{NSE}$, while the composition of the network can be ignored at temperatures greater than $T_\text{high}$.
Specifically, we adopt $T_\text{NSE} = 5.8\units{GK}$, $T_\text{high} = 6.2\units{GK}$, and use the SFHo \ac{EOS} \citep{Steiner2013} for all our simulations.
In case the material does not reach \ac{NSE} (i.e. stays in the Helmholtz regime), the network retains and evolves the initial composition from the progenitor.
We additionally modified the implementation of the hybrid \ac{EOS} \citep{Witt2021} so that the transition follows the same temperature-based criterion as the network, using the same temperature threshold $T_\text{NSE}$.

It should be noted that our reduced network implementation does not include weak reaction rates.
As pointed out by \citet{Navo2023}, this is a reasonable approximation because the simulation time is usually shorter than the half-life of the evolved species.
On the other hand, charged-current neutrino interactions can influence the nucleosynthesis by altering the electron fraction.
In the high-density \ac{EOS} regime, neutrino heating and the effect of charged-current reactions on the electron fraction are both accounted for.
This is not the case in the Helmholtz regime, where neutrino heating is included but the electron fraction is frozen.
However, neutrino reactions are calculated using the opacities given by the \nulib{} table, which is based on the high-density \ac{EOS} table and is therefore not fully consistent with the composition evolved by the network once in the Helmholtz regime.
Because this regime is only entered at relatively low densities and temperatures, where neutrino interactions rapidly become subdominant, we regard this limitation as a minor source of uncertainty in our simulations.

\subsection{Nucleosynthesis post-processing} \label{sec:methods_post_processing}

We compare the results of the internal network with the nucleosynthesis calculations performed with an external network in a post-processing step.
The calculations are performed with the open-source nuclear reaction network \winnet{} \citep{Reichert2023}, including $\sim3000$ nuclear species up to \nucleus{Cerium}{Ce}[195], with nuclear reaction rates taken from the \reaclib{} database \citep[][version of September 19th 2020]{Cyburt2010}.
Neutrino reactions are included only for free nucleons.
Additionally, we performed these calculations using only the 25 species included in the RN16 network or the 242 species included in the RN94 network to validate the implementation of the internal network.

The thermodynamic histories serving as input for the \winnet{} calculations record the time evolution of density, temperature, electron fraction, and neutrino luminosities and average energies along the trajectory of a fluid element in the simulation.
The trajectories are reconstructed from the simulation output by placing tracer particles at the end of the simulation and integrating their paths backward in time from the velocity field using a similar approach as in, e.g. \citet{Witt2021,Sieverding2023b}.
The integration employs a fourth-order Runge-Kutta scheme in time and linear interpolation in space, with an interval of $1\units{ms}$ between simulation outputs.
While \flash{} also provides the capability to evolve passive tracer particles inline with the hydrodynamics \citep{Dubey2012}, the post-processing approach conjointly with backward integration allows us to optimise the number and placement of particles.
For the 2D simulations, we place $10,000$ tracers directly within the unbound material at the end of the runs, where the mass of each tracer is calculated based on the local density and number of tracers.
This results in tracer masses ranging from $10^{-9}$--$10^{-5}\Msun$, recovering the ejected mass to within $1\%$.
For each tracer, the nucleosynthesis calculation begins from the last time the temperature drops below the \ac{NSE} threshold of $5.8\units{GK}$ where the initial composition is given by \ac{NSE}.
If a particle does not reach \ac{NSE}, the calculation starts from the beginning of the simulation using the composition given by the progenitor model instead.
Because we include neutrino interactions, \winnet{} evolves the electron fraction independently of the tracer history, but follows the density and temperature evolution of the trajectory.

\subsection{Progenitor models} \label{sec:methods_progenitors}

In this paper, we present \ac{CCSN} simulations of two progenitor stars evolved in binary systems.
Specifically, and for the remainder of this paper, we refer to the primary, initially more massive star, in the system which is the first to explode.
Both progenitors are obtained from large sets of binary stellar evolution presented in \citet{Ercolino2024} and \citet{Ercolino2025}.
They were evolved self-consistently from the \ac{ZAMS} with the open-source stellar evolution code \mesa{} \citep[version r10398;][]{Paxton2011,Paxton2013,Paxton2015,Paxton2018,Paxton2019} at solar metallicity \citep[$\Zsun = 0.0154$:][]{Asplund2021}, and both underwent phases of stable mass transfer through \ac{RLOF}.

From \citet{Ercolino2024}, we select a \ac{RSG} progenitor with a \ac{ZAMS} mass $\Mzams = 12.6\Msun$, evolved in a binary system with initial mass ratio $q_i = 0.95$ and orbital period $P_i = 1995\units{d}$.
Due to the large orbital separation, this progenitor initiates mass transfer only after core \helium{} exhaustion (Case C; \citealt{Marchant2024}), and until core collapse.
As a result, its final core structure is mostly unaffected by the mass transfer episode and can be regarded as `single-star-like'.
Nonetheless, the mass transfer episode is enough to strip this star of $\sim4\Msun$ of \hydrogen{}-rich material, leaving it with a pre-\ac{SN} mass of $6.78\Msun$ \citep[see][]{Ercolino2024}.
We henceforth refer to this model simply as `\bRSG{}'.

From \citet{Ercolino2025}, we select a \ac{HeS} progenitor with a \ac{ZAMS} mass $\Mzams = 13.8\Msun$ which has been completely stripped of its hydrogen-rich envelope through winds and stable Case B \ac{RLOF}, and is left with a pre-\ac{SN} mass of $3.07\Msun$.
It was evolved in a binary system with initial mass ratio $q_i = 0.5$ and orbital period $P_i = 13.2\units{d}$. We henceforth refer to this model as `\bHeS{}'.

To estimate the impact of binary evolution on the explosion outcome, we compare these models to single-star progenitors with the same initial mass and metallicity.
These progenitors are henceforth referred to as \sRSG{} and \sHeS{} and are compared to the \bRSG{} and \bHeS{} progenitors, respectively.
While the \sHeS{} single-star progenitor was evolved ad-hoc to compare to the \bHeS{} model of \citet{Ercolino2025}, the \sRSG{} progenitor corresponds to a binary model with initial mass ratio $q_i = 0.95$ and orbital period $P_i = 3995\units{d}$ evolved by \citet{Ercolino2024}.
Because this model was evolved in such a wide binary where no \ac{RLOF} occurs, it effectively corresponds to a single-star progenitor.

We summarise all four progenitors in \cref{tab:progenitors}, where we also include the total pre-\ac{SN} mass $M_\text{pre-SN}$ and \carbon[]-\oxygen[] core mass $M_\text{CO~core}$, where the \carbon[]-\oxygen[] core mass corresponds to the mass enclosed at the \ac{CO/He} shell interface, which we define as the location at the bottom of the \helium[] shell where the \helium[] mass fraction drops below half of its maximum value.
We give the central carbon mass fraction at core \helium[] depletion $X_\text{C}$, and also report the location of the \ac{Si/O} composition interface $R_\text{Si/O}$ and compactness parameter \compactness{1.75}, as well as the time of accretion of the \ac{Si/O} interface through the shock, $\tilde{t}_\text{accr}$, and the strength of the density jump at the \ac{Si/O} interface $\delta \rho_\text{Si/O}^2/\rho_\text{Si/O}^2$.
The latter two quantities are defined in \citet{Boccioli2024} and are used to estimate the explodability of each progenitors, discussed in \cref{sec:compare_compactness}.

\begin{table*}
    \caption{
        Overview of the progenitor models.
    }
    \label{tab:progenitors}
\begin{threeparttable} 
    \begin{tabular}{lcccccccccc}
        \hline
        Progenitor & Evolution & $\Mzams$ & $M_\text{pre-SN}$ & $M_\text{CO~core}$ & $X_\text{C}$ & $R_\text{Si/O}$ & $M(R_\text{Si/O})$ & $\tilde{t}_\text{accr}$\tnote{a,*} & $\delta \rho_\text{Si/O}^2/\rho_\text{Si/O}^2$ & \compactness{1.75} \\
                   &           & $[\Msun]$ & $[\Msun]$ & $[\Msun]$ & & $[\mathrm{cm}]$      & $[\Msun]$ & $[\mathrm{s}]$ &         &         \\
        \hline
        \bRSG{}       & Binary    & $12.6$    & $6.78$    & $2.10$    & $0.3729$ & $2.39 \times 10^{8}$ & $1.62$    & $0.341$        & $0.078$ & $0.296$ \\
        \bHeS{}       & Binary    & $13.8$    & $3.07$    & $1.85$    & $0.4247$ & $2.13 \times 10^{8}$ & $1.60$    & $0.279$        & $0.086$ & $0.263$ \\
        \sRSG{}       & Single    & $12.6$    & $10.79$   & $2.11$    & $0.3729$ & $2.69 \times 10^{8}$ & $1.62$    & $0.345$        & $0.088$ & $0.319$ \\
        \sHeS{}       & Single    & $13.8$    & $11.31$   & $2.49$    & $0.3493$ & $2.85 \times 10^{8}$ & $1.64$    & $0.377$        & $0.011$ & $0.353$ \\
        \hline
    \end{tabular}
    \begin{tablenotes}
        \item[a] Time of accretion of the Si/O interface, estimated from the baseline 1D simulations presented in \cref{sec:results}.
        \item[*] All times are given post-bounce.
    \end{tablenotes}
\end{threeparttable}
\end{table*}

\section{Explodability} \label{sec:explodability}

\begin{figure*}
    \includegraphics[width=\textwidth]{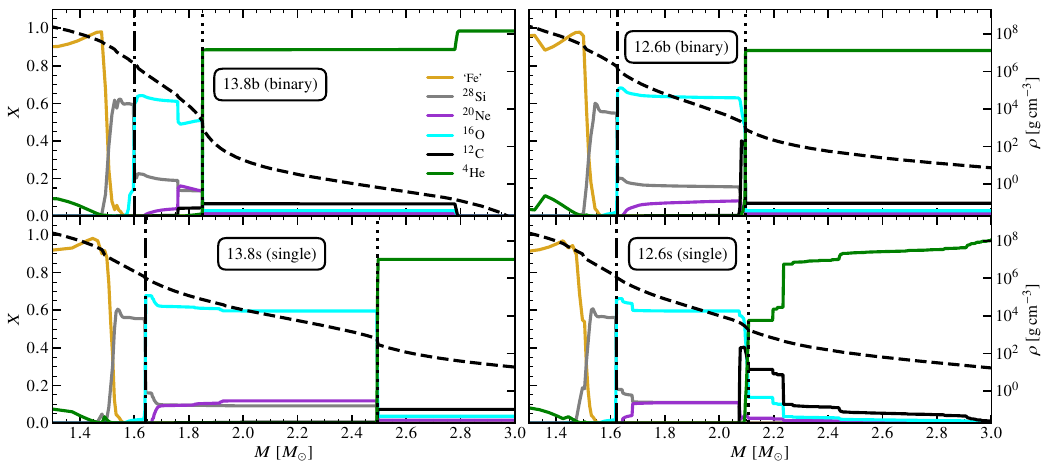}
    \caption{
        Composition (solid lines) and density (black dashed lines) profiles of the \bHeS{} (top left) and \bRSG{} (top right) binary progenitors, and their corresponding \sHeS{} (bottom left) and \sRSG{} (bottom right) single-star progenitors.
        The location of the \ac{Si/O} (vertical black dash-dotted line) and \ac{CO/He} (vertical black dotted line) interfaces are shown for each model.
    }
    \label{fig:core_structure}
\end{figure*}

We start by comparing the pre-\ac{SN} structure of the binary progenitors with that of their analogous single-star progenitors.
\Cref{fig:core_structure} shows the mass fractions as a function of enclosed mass of the most abundant species to which we superimpose the density profiles of each progenitor.
Here, the generic element `Fe' denotes the sum of all \iron{}-peak elements.
The location of the \ac{Si/O} and \ac{CO/He} shell interfaces, given in \cref{tab:progenitors}, are also shown.

\subsection{Compactness} \label{sec:compare_compactness}

We give in \cref{tab:progenitors} the compactness parameter for our different progenitors.
The compactness parameter, first introduced in \citet{OConnor2011}, is defined as
\begin{equation}
    \xi_M = \frac{M / \Msun}{R(M_\text{bary} = M) / 1000\units{km}} \bigg\rvert_{t=t_\text{bounce}} \punct{,}
\end{equation}
where $R(M)$ is the radius that encloses a baryonic mass $M$, and is evaluated at time of bounce.

We evaluate this parameter at $1.75\Msun$, where it encloses the \ac{Si/O} interface while staying within the \carbon[]-\oxygen[] core for all our models.
The compactness parameter characterises the core structure of the collapsing star, and was originally found to strongly correlate with the time of \ac{BH} formation \citep{OConnor2011} in failed \acp{CCSN}.
It has then often been used as an indicator of the final outcome \citep[e.g.][]{Sukhbold2016,Sukhbold2018,Farmer2023}, where low-compactness progenitors are assumed to explode, while high-compactness ones form \ac{BH}.

In their work, \citet{Laplace2021} observed that the compactness of binary stars at the onset of collapse is systematically lower than that of single-star models with the same initial mass.
\citet{Vartanyan2021} then showed that this property tended to correlate with greater explodability of the binary models.
However, \citet{Laplace2021} also showed that this difference disappears when comparing single- and binary-stars of similar \helium[] core mass instead.
When comparing their 3D \ac{CCSN} simulations, \citet{Muller2019} also found no significant evidence of differences in the explosions of single- and binary-star progenitors of comparable \helium[] core mass that could not be associated with stochastic variations already observed in single-star models.
However, as pointed out by \citet{Laplace2021}, there are appreciable differences between the composition profiles of single and binary models, even when comparing stars of similar \helium[] core mass.
The study of binary-evolved stars thus remains essential to nucleosynthesis calculations and \ac{GCE} modelling.

Based on large sets of \ac{CCSN} models, recent studies have argued that the compactness alone does not suffice to predict the explosion \citep[e.g.][]{Couch2020,Vartanyan2023}.
These works found that low-compactness progenitors may fail to explode, while the highest-compactness ones can sometimes produce the most energetic explosions.
Similarly, intermediate-compactness progenitors may more often produce \ac{BH} than high-compactness stars were originally anticipated to do.
\citet{Brown2001,Patton2020,Laplace2025} have argued that the carbon mass fraction in the \carbon[]-\oxygen[] cores also play a crucial role.
In contrast, previous works by \citet{Nakamura2015,Burrows2024} found that the core compactness correlates rather well with other properties of the explosion, such as the explosion energy and \nickel[56] production.

\subsection{Composition interface} \label{sec:compare_core}

The compactness alone can generally not be used to reliably predict if a successful explosion will take place, but may be used together with other quantities to form a more robust explosion criterion.
In particular, recent studies have highlighted the significance of the \ac{Si/O} interface for the explodability \citep[see, e.g.][]{Vartanyan2018,Wang2022,Bruenn2023,Kumar2024,Boccioli2023,Boccioli2025}, and found that a pronounced drop in density near its location can promote the explosion if it is accreted early enough through the shock.
Although few studies have explored the explodability of binary progenitors, current results from self-consistent \ac{CCSN} simulations indicate that binary stars with initial masses in the range $11$--$21\Msun$ explode more readily than their single-star counterparts, and that this behaviour is consistent with deeper, more prominent \ac{Si/O} interfaces \citep{Vartanyan2021}.
However, this trend may reverse for higher mass progenitors, where \citet{Kumar2024} found that binary stars with masses between $37$--$41\Msun$ fail to explode where single stars succeed, which they attribute to their lower compactness compared to single stars.

For our progenitors, we find that the \bHeS{} binary model exhibits a stronger and deeper density drop near the \ac{Si/O} interface than the corresponding \sHeS{} single-star model, as shown in \cref{fig:core_structure}.
This suggests that the \bHeS{} progenitor, whose core has been more strongly affected by binary interaction due to Case B mass transfer, should be more explodable than its single-star equivalent.
On the other hand, the \bRSG{} model shows a core structure overall very similar to that of the \sRSG{} model; their \ac{Si/O} interfaces are located at similar mass coordinates and exhibit comparable density jumps.
This confirms that the late binary interaction has only weakly affected the evolution of the \bRSG{} model, and that this progenitor is in fact single-star like.

Recently, \citet{Kumar2024} proposed a criterion based on the compactness at $2\Msun$ and the properties of the \ac{Si/O} shell interface.
They derived this criterion from a systematic study of 1D explosion models of single and binary stars performed with the \grid{} code \citep{OConnor2010} and using the \stir{} method \citep{Couch2020,Boccioli2021}.
According to this criterion, there will be an explosion if the \ac{Si/O} shell interface is accreted by the shock within $300\units{ms}$ after bounce ($\tilde{t}_\text{accr} < 0.3\units{s}$) and if the density jump at the interface satisfies $\delta \rho_\text{Si/O}^2/\rho_\text{Si/O}^2 > 0.08$, where $\rho_\text{Si/O}$ is the density interior to the interface and $\delta \rho_\text{Si/O}$ is the jump across the interface \citep[for details, see][]{Boccioli2023,Kumar2024}.
The corresponding values for the different progenitors are shown in \cref{tab:progenitors}.
The \bHeS{}, \bRSG{}, and \sRSG{} progenitors all show a strong enough density jump, but the \ac{Si/O} interface is accreted too late by all but the \bHeS{} model.
Thus, our \bHeS{} binary progenitor can be expected to be the most explodable of our models.

Alternatively, \citet{Maltsev2025} developed a set of explodability criteria based on pre-\ac{SN} stellar structure variables: the compactness \compactness{2.5}, the central specific entropy $s_c$, and the \citet{Ertl2016} parameters $\mu_4$ and $\mu_4 M_4$.
They placed lower limits on these quantities below which the explosion would be predicted to succeed, and upper limits above which the explosion would fail.
In the region within these limits, where the final fate is ambiguous, they identified a separation line in the ($\mu_4 M_4$, $\mu_4$) plane above which the explosion is predicted to succeed.
According to these criteria, all our progenitors should explode as their compactness are below the lower threshold of $0.314$ \citep[see][table 2]{Maltsev2025}.

\section{Results} \label{sec:results}

We performed 1D and 2D \ac{SN} explosion simulations based on the single and binary progenitors presented in \cref{sec:methods_progenitors}.
An overview of all explosion models presented in this work is given in \cref{tab:simulation_overview}, where we include the final baryonic \ac{PNS} mass $M_\text{PNS}$, unbound mass $M_\text{ej}$, diagnostic explosion energy $E_\text{exp}$ \citep[as defined in][]{OConnor2018a}, and total integrated contribution to the explosion energy from nuclear reactions at the end of the simulations $Q_{nuc}$.
Each model is given a descriptive name constructed from the name of the progenitor, the network used (if any), and the dimensionality of the simulation.
For each progenitor shown in \cref{tab:progenitors} we ran a model without network in 1D and 2D, which constitute the baseline for our comparison between single and binary-star progenitors in \cref{sec:results_progenitors}.
We use the networks only for the binaries where we consider two variants for each network depending on whether feedback from nuclear energy generation is enabled (marked as `RN16e' and `RN94e') or not (marked as `RN16' and `RN94').

\begin{table*}
    \caption{
        Overview of the explosion models.
    }
    \label{tab:simulation_overview}
\begin{threeparttable}
    \begin{tabular}{lcccccccccc}
        \hline
         Model & Dimensionality & Network & $t_\text{exp}$\tnote{a,*} & $t_\text{end}$\tnote{*} & $M_\text{PNS}$ & $M_\text{ej}$\tnote{b} & $E_\text{exp}$ & $M({}^{56}\text{Ni})$ & $Q_\text{nuc}$ \\
               &                &         & [s]                                 & [s]                               & [$\Msun$]      & $[\Msun]$     & [$\times 10^{50}$ erg]          & [$\times 10^{-2} \Msun$]     & [$\times 10^{50}$ erg]          \\
        \hline
         \bRSG{}\_1D        & 1 & None & $0.336$ & $1.385$ & $1.60$ & $0.331$ & $3.8$ & Not Included & Not included          \\
         \bRSG{}\_RN16\_1D  & 1 & RN16 & $0.336$ & $1.385$ & $1.60$ & $0.331$ & $3.8$ & $3.98$ & Not included          \\
         \bRSG{}\_RN16e\_1D & 1 & RN16 & $0.316$ & $1.385$ & $1.59$ & $0.336$ & $4.1$ & $4.66$ & $0.78$ \\
         \bRSG{}\_RN94\_1D  & 1 & RN94 & $0.336$ & $1.385$ & $1.60$ & $0.331$ & $3.8$ & $3.53$ & Not included          \\
         \bRSG{}\_RN94e\_1D & 1 & RN94 & $0.315$ & $1.385$ & $1.59$ & $0.336$ & $4.1$ & $4.25$ & $0.78$ \\
         \bRSG{}\_2D        & 2 & None & $0.397$ & $1.240$ & $1.68$ & $0.078$ & $2.0$ & Not Included & Not included          \\
         \bRSG{}\_RN16e\_2D & 2 & RN16 & $0.394$ & $1.230$ & $1.67$ & $0.061$ & $1.3$ & $0.53$ & $0.19$ \\
         \hline
         \bHeS{}\_1D        & 1 & None & $0.292$ & $1.384$ & $1.59$ & $0.248$ & $3.2$ & Not Included & Not included          \\
         \bHeS{}\_RN16\_1D  & 1 & RN16 & $0.292$ & $1.384$ & $1.59$ & $0.248$ & $3.2$ & $4.73$ & Not included          \\
         \bHeS{}\_RN16e\_1D & 1 & RN16 & $0.290$ & $1.384$ & $1.59$ & $0.255$ & $4.0$ & $4.92$ & $0.83$ \\
         \bHeS{}\_RN94\_1D  & 1 & RN94 & $0.292$ & $1.384$ & $1.59$ & $0.248$ & $3.2$ & $4.39$ & Not included          \\
         \bHeS{}\_RN94e\_1D & 1 & RN94 & $0.295$ & $1.384$ & $1.59$ & $0.256$ & $4.2$ & $4.26$ & $0.77$ \\
         \bHeS{}\_2D        & 2 & None & $0.286$ & $1.284$ & $1.66$ & $0.080$ & $1.5$ & Not Included & Not included          \\
         \bHeS{}\_RN16e\_2D & 2 & RN16 & $0.362$ & $1.263$ & $1.66$ & $0.074$ & $1.4$ & $0.68$ & $0.22$ \\
         \hline
         \sRSG{}\_1D        & 1 & None & $0.308$ & $1.400$ & $1.58$ & $0.392$ & $3.9$ & Not Included & Not included          \\
         \sRSG{}\_2D        & 2 & None & $0.385$ & $1.203$ & $1.67$ & $0.059$ & $1.4$ & Not Included & Not included          \\
         \hline
         \sHeS{}\_1D        & 1 & None & $0.270$ & $1.403$ & $1.57$ & $0.430$ & $3.7$ & Not Included & Not included          \\
         \sHeS{}\_2D        & 2 & None & $0.256$ & $1.185$ & $1.66$ & $0.101$ & $1.8$ & Not Included & Not included          \\
        \hline
    \end{tabular}
    \begin{tablenotes}
        \item[a] Time when the average shock radius reaches $400\units{km}$.
        \item[b] Total unbound mass at the end of the simulation.
        \item[*] All times are given post-bounce.
    \end{tablenotes}
\end{threeparttable}
\end{table*}

Note that we only used the small network variant with nuclear energy generation enabled (RN16e) for the 2D simulations.
This decision is informed by the results of the 1D simulations, but also motivated by the non-negligible computational cost incurred by the large network in multidimensional simulations.

\subsection{Comparison between single and binary stars} \label{sec:results_progenitors}

\begin{figure}
    \includegraphics[width=\columnwidth]{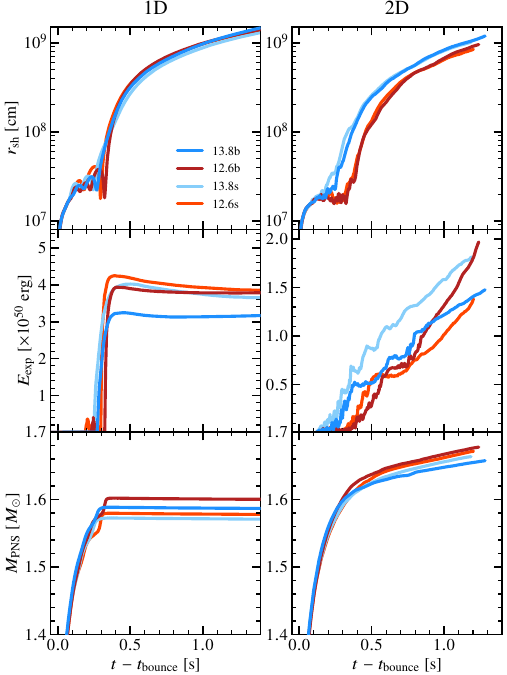}
    \caption{
        Time evolution of the shock radius (top), diagnostic explosion energy (middle), and \ac{PNS} mass (bottom) for the different baseline models in 1D (left) and 2D (right).
    }
    \label{fig:base_explosion_1d_2d}
\end{figure}

\Cref{fig:base_explosion_1d_2d} shows the evolution of the shock radius $r_\text{sh}$ (average shock radius in 2D), the diagnostic explosion energy $E_\text{exp}$, and the baryonic \ac{PNS} mass $M_\text{PNS}$ for the baseline 1D and 2D explosion models.

As discussed in \cref{sec:explodability}, only our \bHeS{} progenitor is predicted to explode according to the explodability criterion of \citet{Boccioli2023}, but our 2D simulations show that all of them explode\footnote{We do not consider the 1D simulations here as the heating factor is tuned to produce explosions.}.
Overall, the \bRSG{} and \sRSG{} models produced similar explosions: they explode around the same time and result in very similar explosion energies, both in 1D and 2D.
This is again reassuring as both stars share similar core structures.
Furthermore, the runaway shock expansion is triggered around the same time as the \ac{Si/O} shell interface is accreted in all but the \sHeS{} model.
Interestingly, this last model explodes earlier than the others despite being presumably less explodable, placing it as a potential outlier.
Although this behaviour should be studied more systematically in multidimensional simulations using a larger set of progenitors, our simulations agree with previous results from, e.g. \citet{Vartanyan2018,Vartanyan2021,Boccioli2023,Boccioli2024}, and seem to reinforce the connection between the \ac{Si/O} shell interface and explodability.

In terms of the explosion energy, the \bHeS{} model produces the least energetic explosion with a final value of $0.32\units{B}$ in 1D.
The \bRSG{} and \sRSG{} models are slightly more energetic with explosion energies of $\sim0.38\units{B}$ and $\sim0.39\units{B}$, respectively, in 1D.
The \bHeS{} model is also less energetic than the \bRSG{} in 2D, with energies of $0.15\units{B}$ and $0.2\units{B}$, respectively.

For the \bHeS{}, \bRSG{}, and \sRSG{} models, we recover a positive correlation between compactness and explosion energy, which was previously observed by, e.g. \citet{Burrows2024}, while the \sHeS{} model stands out again as a potential outlier from the 1D results.
Quantitatively, the explosion energies are also in agreement with \citet[figure~12]{Burrows2024} when compared to the compactness, albeit slightly under-energetic at the end of the 2D simulations.
Finally, our results agree with the (very rough) correlation between the compactness and the post-bounce explosion time seen in \citet[figure~5]{Burrows2024}
It should, however, be noted that the explosion energies are not converged but still rapidly rising at the end of the 2D simulations, at $\sim1.2\units{s}$ post-bounce, as is usually observed in 2D models \citep[see e.g.][]{Muller2015,Bollig2021,Witt2021}.

For comparison, we also computed the explosion properties predicted by the semi-analytic model of \citet{Muller2016}, using the parameter calibration of \citet{Schneider2021}, for our four progenitors.
This method predicts explosion energies of $0.9$--$1.3\units{B}$ for our progenitors, systematically higher than our 1D results by a factor of $\sim2$--$3$ across all our models.
Semi-analytic, population-calibrated explosion models of this kind are widely used to derive explosion properties for large grids of stellar progenitors \citep[e.g.][]{Schneider2021,Gilkis2025,Ercolino2026}, but tensions between their predictions and those of self-consistent explosion simulations have been noted before \citep[e.g.][]{Vartanyan2021}.
This comparison is, however, subject to important limitations.
In 1D, our heating factor of $2.8$ was chosen only to ensure that all progenitors explode, rather than to reproduce a specific target explosion-energy distribution as in the population-calibrated model of \citet{Schneider2021}.
In 2D, our explosion energies are not yet converged and are still rising at $\sim1\units{s}$ post-bounce.
At this stage, our 2D models reach explosion energies of $\sim0.14$--$0.2\units{B}$, similar to the typical $\sim0.2\units{B}$ diagnostic explosion energy reported by \citet{Vartanyan2021} for their exploding models within the first second after bounce.
Furthermore, our results correspond to the diagnostic explosion energies, which must be corrected for the gravitational overburden of the envelope.
As emphasised by \citet{Vartanyan2021}, resolution of these  discrepancies requires longer, self-consistent 3D simulations.
We leave a more systematic assessment of this offset, across a larger progenitor set, to future work.

\subsection{Nucleosynthesis} \label{sec:results_nucleosynthesis}

Large nuclear networks are necessary to capture the full range of nucleosynthetic yields produced in \acp{CCSN} and to accurately determine the energy released by nuclear reactions.
Both the change in composition and the feedback from nuclear energy generation can affect the dynamics of the explosion, however, the inclusion of such networks in \ac{CCSN} simulations is heavily limited by the computational cost incurred.
In this work, we show that a small $\alpha$-chain nuclear reaction network provides a good approximation to nuclear energy generation and the production of key species, particularly \nickel[56] \citep{Timmes2000a,Utrobin2015,Harris2017,Navo2023}.

\subsubsection{Composition} \label{sec:networks_nucleosynthesis}

We first discuss the nucleosynthesis results from the models \bHeS{}\_RN16e\_1D, \bHeS{}\_RN94e\_1D, and \bHeS{}\_RN16e\_2D, including post-processing with \winnet{} (see \cref{tab:simulation_overview}).
In 1D, we do not consider the two RN16 and RN94 variants (with nuclear energy generation disabled) as the nucleosynthesis is overall similar to that of the corresponding models with nuclear energy generation enabled.
For the remainder of this section, we refer to the ejecta as the unbound, shocked material in the simulations.
This corresponds to the total mass contributing to the diagnostic explosion energy at any time.

\begin{figure}
    \includegraphics[width=\columnwidth]{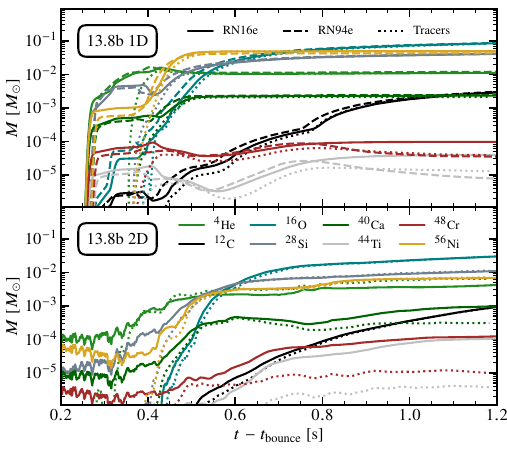}
    \caption{
        Time evolution of the composition of the ejecta from the internal networks and post-processed tracers for the \bHeS{} explosion models.
        The curves show the total unbound mass of selected key isotopes as predicted by the RN16e (solid lines) and RN94e (dashed lines) network variants, and the post-processed tracers (dotted lines).
        The top panel shows the 1D models, while the bottom panel shows the 2D model which only includes the RN16e and tracer calculations.
    }
    \label{fig:ejecta_evolution}
\end{figure}

The evolution of the composition of the ejecta given by the different in-situ networks and tracer calculations is shown in \cref{fig:ejecta_evolution} for few key isotopes.
Results from the internal networks and post-processed tracers are generally in good agreement in both 1D and 2D.
At about $400\units{ms}$, the shock front cools below the \ac{NSE} temperature threshold of $5.8\units{GK}$.
\nickel[56] is then produced explosively and rapidly converges to its final abundance about $100\units{ms}$ later, when the shocked material cools below $\sim 4\units{GK}$.
The abundances of other $\alpha$-elements such as \carbon[12], \oxygen[16], and \silicon[28] continue to increase for a longer time as the shock crosses the \silicon[]- and \oxygen[]-rich shells, the latter containing also traces of \carbon[12].
The evolution of the composition of the ejecta in the 2D models (bottom panel of \cref{fig:ejecta_evolution}) is qualitatively similar to that of the 1D models.
However, at $t_\text{pb} = 1.2\units{s}$, the total unbound mass in the 2D model is about five times smaller than in the 1D models.
Consequently, the total masses of most species are diminished by a similar factor.

When comparing in-situ with post-processing, the only clear differences appear in the evolution of \calcium[40], \titanium[44], and \chromium[48].
In 1D, the large network and the post-processed tracers are in good agreement for these species, whereas the small network overproduces \titanium[44] and \chromium[48] by factors of $5$ and $3$, respectively.
These discrepancies exceed an order of magnitude in the 2D results, where \calcium[40] is also overproduced by a factor of $3$ in the small network.
In particular, the destruction of \titanium[44] by proton capture becomes dominant over its production once the shock front cools below $\sim2\units{GK}$, which is clearly visible at $\sim800\units{ms}$ in the 1D model (top panel of \cref{fig:ejecta_evolution}).
A similar behaviour is seen for \chromium[48].
This corresponds to the `($p$, $\gamma$)-leakage regime', in which the nuclear flow leaks from \titanium[44] to heavier nuclei through successive proton captures before decaying back towards stability \citep{Magkotsios2010,Magkotsios2011}.
The RN16 network is too small to capture this process, leading to a significant overestimation of \titanium[44] and \chromium[48].

\begin{figure}
    \includegraphics[width=\columnwidth]{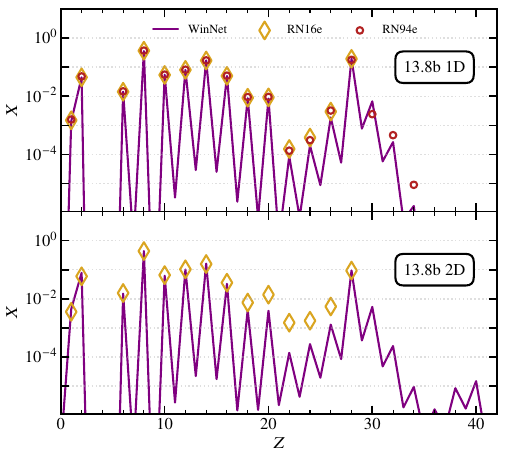}
    \caption{
        Composition of the ejecta at the end of the different explosion models of the \bHeS{} progenitor.
        The 1D models (top panel) include results from the RN16e (golden diamonds) and RN94e (red circles) network variants, while only the RN16e network is available in 2D (bottom panel).
        The detailed composition of the ejecta is calculated in post-processing with \winnet{} using tracer particles obtained from the RN16e 1D and 2D simulations (purple lines).
    }
    \label{fig:ejecta_yields}
\end{figure}

\Cref{fig:ejecta_yields} shows that the elemental abundances at the end of the different 1D and 2D models are in good agreement with those calculated in post-processing with \winnet{}.
Once again, the largest discrepancies are found for \titanium[], \chromium[], and \iron[], as well as \argon[] and \calcium[] in the 2D case.
In contrast, the abundances of the other $\alpha$ elements common to each network are in almost complete agreement across all models and post-processing calculations.
From the ejecta, we find that the material that reaches \ac{NSE} contributes to \helium[4] and most of the elements with $Z \gtrsim 20$.
Notably, the recombination of $\alpha$ particles with heavy nuclei from \ac{NSE} produces the majority of the ejected \nickel[] by the time the shock front has cooled to $\sim4\units{GK}$.
On the other hand, lighter elements with $Z \lesssim 20$ originate primarily from the material that is shocked without reaching \ac{NSE}.
Of this part of the ejecta, we find that $5-10\%$ of the \carbon[], \oxygen[], \neon[], and \magnesium[] initially present in the progenitor is destroyed after crossing the shock front by the end of the simulation.
These elements are burned primarily into \silicon[], \argon[], and \calcium[], but produce traces of heavier elements up to the \iron[]-peak as well.

\begin{figure*}
    \includegraphics[width=\textwidth]{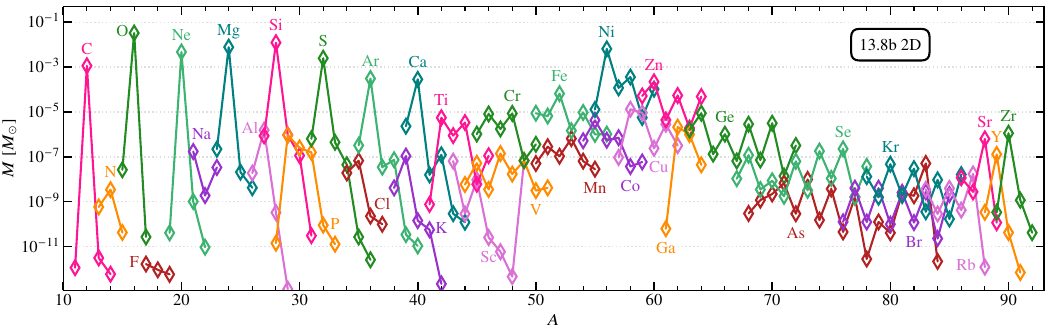}
    \caption{
        Isotopic composition of the ejecta for elements from \carbon[] to \nucleus{zirconium}{Zr}[] as predicted by the post-processed tracers at the end of the \bHeS{}\_RN16e\_2D simulation. Every line connects different isotopes of the same element.
        Only the most abundant species are shown for clarity.
    }
    \label{fig:isotopic_chains}
\end{figure*}

Elements from \carbon[] to \silicon[] primarily originate from hydrostatic burning, and are found in the upper layers of the \carbon[]\oxygen[] core, which do not reach \ac{NSE} during the explosion.
These layers are characterised by $Y_e \simeq 0.5$, and it can be seen in \cref{fig:isotopic_chains} that elements up to $Z = 20$ are mainly found as their symmetric isotopes ($Z = N$) as a result.
This can thus be well captured by our small network.
This is also a good approximation for \nickel[], which is primarily synthesised as \nickel[56].
However, in 2D, up to $5\%$ of the total amount of \nickel[] is found as \nickel[58], and $\sim1\%$ as \nickel[57] and \nickel[60] reflecting slightly more neutron-rich conditions at freeze-out.
Therefore, the inclusion of \nickel[56] in the small network is generally enough to capture the overall \nickel[] production during the explosion.
In contrast, \cref{fig:isotopic_chains} shows that a significant fraction of the \titanium[] abundance is present as \titanium[42] rather than \titanium[44].
Similarly, the \chromium[] abundance is spread mostly between isotopes from \chromium[45] to \chromium[48], and the \iron[] abundance between \iron[50], \iron[51], \iron[52], and \iron[54].

\begin{figure}
    \includegraphics[width=\columnwidth]{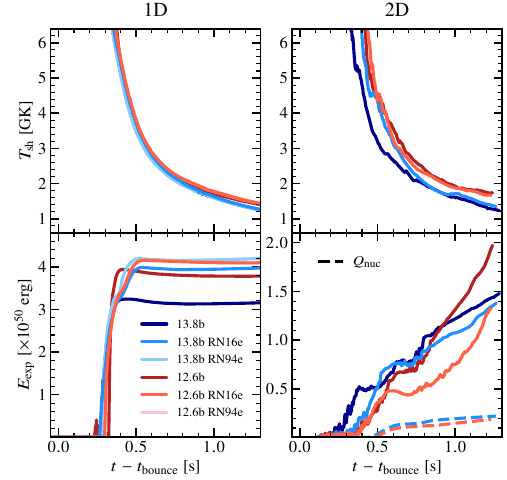}
    \caption{
        Time evolution of the post-shock temperature (top) and diagnostic explosion energy (bottom) for the baseline, RN16e, and RN94e models in 1D (left) and 2D (right).
        The post-shock temperature for the 2D models corresponds to the average temperature behind the front shock.
        The contribution from the total nuclear energy release to the explosion energy is shown for the 2D RN16e simulations (dashed lines).
        The 1D models for the network variants that do not include nuclear energy generation present the same evolution as their corresponding baseline model and are not shown here.
    }
    \label{fig:net_explosion_1d_2d}
\end{figure}

We note that by the end of our simulations, the shock has not yet crossed most of the \oxygen- and \carbon[]-rich material in the progenitor; therefore, we can conjecture that the \ac{SN} explosion only has a minimal impact on the total amounts of \carbon[], \oxygen[], and \neon[] present in the progenitor.
\citet{Farmer2021} showed that this was particularly accurate in the case of \carbon[12], whose total yields resulting from the \ac{CCSN} were largely insensitive to the network size and explosive burning.
This is explained by the fact that the bulk of the \carbon[12] is located too far from the centre of the star so that the shock has cooled too much to efficiently burn it by the time it reaches those layers.
The evolution of the shock temperature in our 1D and 2D models is shown in the top panel of \cref{fig:net_explosion_1d_2d}.
For example, the shock front in the \bRSG{} model does not reach the \carbon{}-rich layer until $\sim1\units{s}$ in 1D and $\sim1.2\units{s}$ in 2D, near the end of the simulation, at which points the shock temperature is below $\sim2\units{GK}$.
This was also verified by \citet{Farmer2023} for most elements up to \magnesium{}, whose final yields post-\ac{SN} were dominated by uncertainties in the progenitor rather than the details of the explosion.

\begin{figure}
    \includegraphics[width=\columnwidth]{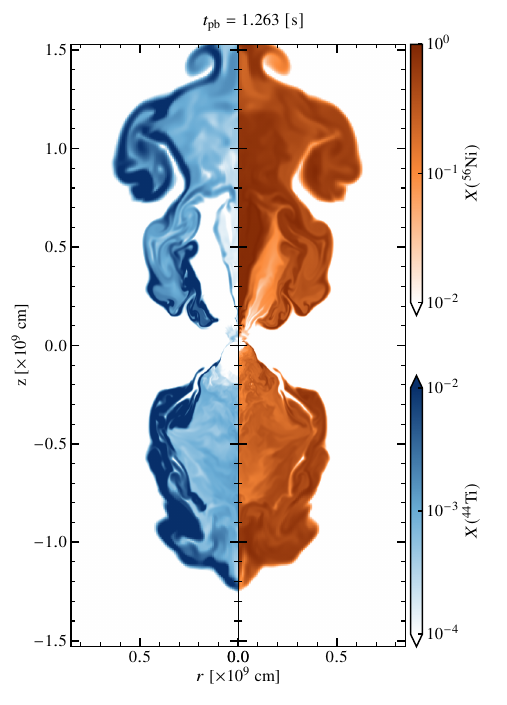}
    \caption{
        Distribution of \titanium[44] (blue, left panel) and \nickel[56] (orange, right panel) at the end of the \bHeS{}\_RN16e\_2D explosion model.
    }
    \label{fig:titanium_distribution}
\end{figure}

The bottom panel of \cref{fig:ejecta_evolution} also shows that \titanium[44] is produced in larger amount and continues to grow towards the end of the 2D simulation, whereas it had mostly converged in the 1D case.
This is consistent with recent results from \citet{Sieverding2023a,Wang2024}, who showed that long-term multidimensional simulations of neutrino-driven \acp{CCSN} were able to produce greater amounts of \titanium[44] than spherically symmetric models.
In these works, the nucleosynthesis was calculated using post-processed, backwardly integrated tracer particles and a large nuclear network, and their results highlighted the importance of non-monotonic thermodynamic histories on the production of \titanium[44].
\Cref{fig:titanium_distribution} shows the spatial distribution of \titanium[44] and \nickel[56] at the end of the \bHeS{}\_RN16e\_2D simulation.
\nickel[56] is quite uniformly distributed, while \titanium[44] is more strongly concentrated around shocks and regions with high entropy gradients, highlighting its sensitivity to the thermodynamic history.

The total mass of \titanium[44] at the end of the \bHeS{}\_RN16e\_2D simulation is $\sim10^{-4}\Msun$, about twice as much as in 1D, but is overestimated by over an order of magnitude compared to the post-processing results, as can be seen in \cref{fig:ejecta_yields}.
It should also be noted that the abundance of \titanium[44] has not yet reached an asymptotic value, thus a longer simulation time would be required to determine the final yield.
Regardless, we do not expect the \bHeS{} progenitor to produce as much as $10^{-4}\Msun$ of \titanium[44] in post-processing, as \citet{Wang2024} were only able to reach this amount for progenitors with $\Mzams \gtrsim 16\Msun$.

Similar differences affecting notably \titanium[], \chromium[], and \iron[] between in-situ nuclear networks and post-processed tracer calculations have been reported in multidimensional simulations using larger internal networks by, e.g. \citet{Harris2017,Navo2023}.
As highlighted in these works, Lagrangian tracer particles less effectively track low-density ejecta, where these isotopes are abundantly produced due to $\alpha$-rich freeze-out.
A much larger number of particles may thus be required to fully resolve these regions.
In addition, these few elements are more sensitive to the thermodynamic histories, and insufficient time resolution of the particles trajectories can lead to significant errors.
In contrast, the results from the in-situ network may be influenced by the grid resolution and numerical diffusion in the simulation, but more importantly by the size of the network.
Consequently, the predictions from our small $\alpha$-chain network for \titanium[], \chromium[], and \iron[] cannot be trusted.
While greater time and spatial resolution of tracer particles should produce more accurate results, finer grids and a larger in-situ network are ultimately needed to address these issues and correctly predict the abundances of these few species.

\subsubsection{Energy generation} \label{sec:networks_energy}

We turn to the impact of nuclear energy generation on the dynamics of the explosion.
Our 1D models show that evolving the composition with a nuclear network, but without accounting for the nuclear energy release, produces no significant difference in the dynamics compared to our baseline models.
Any differences in the dynamics of our simulations thus result from nuclear energy generation rather than changes in the composition \citep[but see][for a minor impact on neutrino absorption]{Navo2023}, for which we distinguish two regimes that can alter the explosion.

\begin{figure}
    \includegraphics[width=\columnwidth]{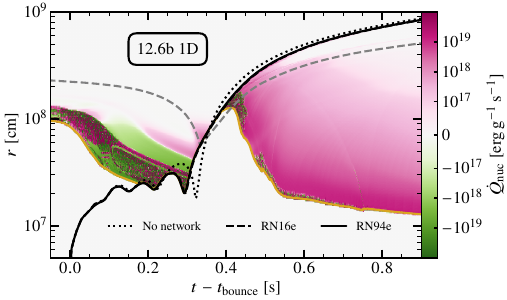}
    \caption{
        Evolution of the specific nuclear energy generation rate $\dot{Q}_\text{nuc}$ in the \bRSG{}\_RN94e\_1D explosion model.
        Pinker hues indicates a net release of energy due to nuclear burning, while greener hues shows that energy is consumed (i.e. dissociation).
        The evolution of the shock radius is nearly the same for the \bRSG{}\_RN16e\_1D (black dashed line) and \bRSG{}\_RN94e\_1D (black solid line) models, while all models that do not include nuclear energy generation show the same behaviour as the baseline \bRSG{}\_1D model (black dotted line).
        The location of the \ac{Si/O} shell interface (grey dashed line) and the $T_\text{NSE}$ threshold, above which $\dot{Q}_\text{nuc}$ is not available, (golden solid line) are also shown.
    }
    \label{fig:engen_1d}
\end{figure}

In the first regime, energy generation emerges from nuclear burning in the unshocked \silicon[] and \oxygen[] shells.
\Cref{fig:engen_1d} shows the evolution of the nuclear energy generation rate $\dot{Q}_\text{nuc}$ across the simulation domain of the \bRSG{}\_RN94e\_1D model.
The $5.8\units{GK}$ \ac{NSE} threshold is indicated by a golden line, and the \ac{Si/O} shell interface by a dashed grey line.
The energy generation rate is nearly identical for the small and large networks and leads to a somewhat earlier shock revival by $\sim30\units{ms}$ compared to the baseline model (i.e. without nuclear energy generation), but also a slower shock expansion afterwards.
Strong variations at the bottom of the \silicon[] shell, just above the \ac{NSE} threshold, indicate that the nuclear reactions are in near equilibrium, i.e. \ac{NSE}.
Two distinct layers can also be seen: one exhibits a net cooling due to nuclear reactions, while the other, closer to the \ac{Si/O} interface, shows a net heating.
This is explained by the rise in temperature during collapse, which triggers \silicon[] burning and releases energy.
However, when the material reaches $\sim4\units{GK}$, it enters a \ac{QSE} that rapidly re-adjusts the composition and consumes energy instead.
This behaviour may affect shock revival when the material is accreted by the stalling shock; however, it may depend upon the composition given by the progenitor model, and be influenced by the use of different nuclear networks and \ac{EOS} between the stellar evolution and \ac{SN} codes.
We also note that the shock dynamics for the \bHeS{} models is nearly unaffected by nuclear energy in 1D, while shock revival is delayed by $\sim100\units{ms}$ relative to the baseline in 2D.
While we do not find a clear systematic impact of nuclear burning on shock revival, further investigation is required, in particular with a more consistent pipeline connecting the late stages of stellar evolution to the \ac{SN} simulations \citep[e.g.][]{Wongwathanarat2015,Muller2017,Bollig2021,Vartanyan2025,Andresen2026}.
Lastly, we note a net heating at the bottom of the collapsing \oxygen[] shell, which is accreted by the shock slightly after revival and may facilitate its expansion.

The second regime takes place after shock revival, in the post-shock region.
This is seen in \cref{fig:engen_1d} after $\sim400\units{ms}$, when the shock reaches $5.8\units{GK}$, where a significant amount of energy is released mainly through recombination and continues to power the explosion as the shock expands.
At the end of the simulation, $\sim20\%$ of the total explosion energy is due to this release of nuclear energy, as shown in \cref{tab:simulation_overview}.
Similar values are found for both the small and large networks, although this fraction amounts to $\sim15\%$ of the explosion energy in 2D.
This is overall consistent with the results of \citet{Bruenn2016,Navo2023}, who found a similar range of $10$--$20\%$ across their models.
However, the impact of nuclear energy in the 1D models becomes rapidly negligible due to the very low densities in the neutrino wind.
The explosion energy and the total nuclear energy released in the 1D models both reach asymptotic values within $100$--$200\units{ms}$ after the shock has been revived, as can be seen in the bottom part of \cref{fig:net_explosion_1d_2d}.

\begin{figure*}
    \includegraphics[width=\textwidth]{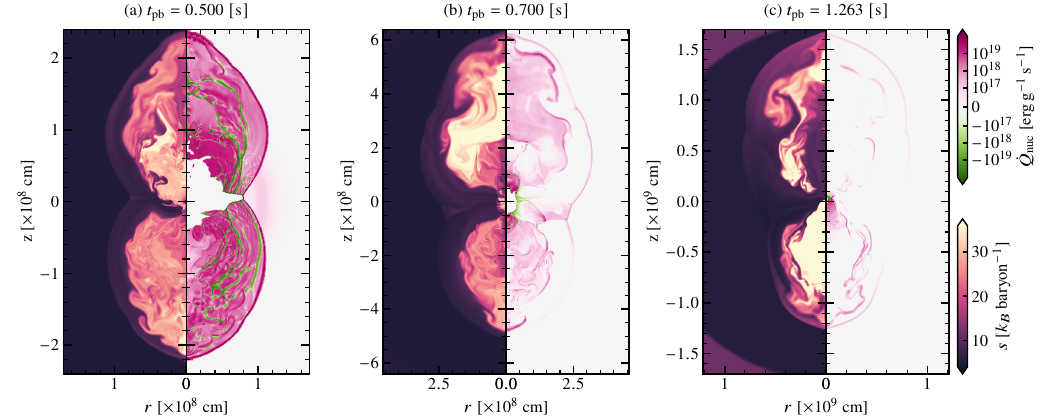}
    \caption{
        Slices of entropy (left panels) and nuclear energy generation rate (right panels) in the \bHeS{}\_RN16e\_2D model at $500\units{ms}$ (a), $700\units{ms}$ (b) and $1263\units{ms}$ (c) after bounce.
        Nuclear energy generation is missing in the central region, close to the \ac{PNS}, because it is in \ac{NSE}.
    }
    \label{fig:engen_2d}
\end{figure*}

This is not true for the 2D models, in which the explosion energy and, to a lesser extent, the contribution from nuclear energy generation are still rising at the end of the simulation.
Nevertheless, the effect of nuclear energy rapidly weakens, as shown by \cref{fig:engen_2d}.
At $500\units{ms}$ (\cref{fig:engen_2d}(a)), the shocked material exits \ac{NSE} and releases a large amount of energy through nuclear reactions.
Then, at $700\units{ms}$ (\cref{fig:engen_2d}(b)), the rate of nuclear energy generation has already dropped by several orders of magnitude and only occurs in regions where the entropy is high.
Finally, by the end of the simulation, at $1263\units{ms}$ (\cref{fig:engen_2d}(c)), nuclear reactions have almost completely ceased.
As discussed above, almost no explosive burning arises as the shock crosses the \oxygen[]-rich shell, however, it has been proposed that explosive \oxygen[]-shell burning may more significantly contribute to nuclear energy generation in very energetic explosions \citep{Bruenn2016}, which we do not observe in our weakly exploding models.

\section{Summary and Conclusions} \label{sec:outro}

Most core collapse supernovae are thought to occur in stars which which have been part in at least one binary mass transfer event during their evolution \citep{Ercolino2026}.
Here, we have compared the pre-\ac{SN} core structure and explosion properties of two binary-evolved progenitor stars with those of single-star progenitors with the same \ac{ZAMS} mass.
The close-binary progenitor \bHeS{}, most affect by binary mass transfer, shows significant differences in its final core structure compared to its single-star counterpart, notably in terms of its compactness parameter \compactness{1.75}, chemical composition, density profile, and the location of the composition interfaces.
On the other hand, the wide-binary progenitor \bRSG{} shows similar properties as the corresponding single-star \sRSG{} progenitor, making it single-star like, owing to its \ac{RLOF} occuring only after core helium exhaustion.

The differences in the pre-\ac{SN} core of the \bHeS{} model ultimately
result in differences in the explosions.
However, we showed that the general explosion properties of our binary progenitors compare well with previous work involving single-star models.
The differences we observe in core properties are consistent with \citet{Laplace2021,Vartanyan2021}, who found that binary progenitors typically exhibit lower core compactness and deeper, more pronounced \ac{Si/O} interfaces than their single-star counterparts of the same initial mass.
In contrast, they found similar values of the compactness when comparing binary and single-star models of same \helium[] core mass at core \helium[] depletion instead.
As the properties of the explosion tend to correlate with the compactness, these results suggest that binary and single-star progenitors can still produce similar explosions for the same value of the compactness, even though the compactness itself is not a good metric of explodability.
Rather, the outcome also depends on the evolution history of the star and cannot be well correlated with the \ac{ZAMS} mass.
Furthermore, \citet{Brown2001,Laplace2021} stressed that the core composition of binary stars presents systematic differences compared to single stars, even for the same reference \helium[] core mass and compactness parameter.
This makes it necessary to explore the explosion of binary-evolved progenitors more systematically in order to produce accurate predictions for \ac{GCE} and observations.

We implemented the 16-species and 94-species nuclear reaction networks from \citet{Navo2023} in the \flash{} code and investigated the impact of nuclear heating on the dynamics of the \ac{SN} explosion.
We compared the nucleosynthesis results from the in-situ networks with post-processing using a large network and tracer particles.
We showed that the use of very large and computationally expensive nuclear reaction networks coupled to the hydrodynamics is generally unnecessary.
Our simulations demonstrate that the bulk of the nucleosynthesis, including \nickel[], can be accurately followed using a small $\alpha$-chain network.
This approximation also holds for $\alpha$-elements up to the \silicon[] group, which essentially originate from the progenitor composition and are mostly unaffected by explosive burning.
The situation is more delicate for \titanium[] and \chromium[], which are still subject to substantial uncertainties, notably due to missing reaction pathways in the in-situ network.
The production of these elements can instead be captured in post-processing using a complete nuclear network; however, this method may still suffer from uncertainties related to the integration of Lagrangian tracer particles.

We also showed that the rate of nuclear energy generation is well approximated by an $\alpha$-chain network.
The RN16 network is nearly complete for \carbon[] and \oxygen[] burning, which may be enhanced during the collapse of the \oxygen[] shell and facilitate shock expansion at the early stage of the explosion.
In addition, the approximation of steady-state species allows the inclusion of additional ($\alpha$, $p$), ($p$, $\gamma$) reaction chains, which can more strongly contribute to the nucleosynthesis and energy generation than ($\alpha$, $\gamma$) reactions for certain species during the explosion.
We see intense nuclear burning in the \silicon[] shell during collapse, but do not find a clear, systematic impact of nuclear energy generation on shock revival which cannot be attributed to stochastic variations or uncertainties in the progenitor composition itself.
This effect should be investigated further using consistent physics between the late stages of the progenitor evolution and its subsequent core-collapse in multiple dimensions.

\section*{Acknowledgements}
We thank Haakon Andresen, Luca Boccioli, Oliver Eggenberger Andersen, Finia Jost, Marta Molero, Gerard Navò, and Moritz Reichert for useful discussions.
Furthermore, we thank Sean Couch for \flash{} development.
This work was supported in part by the Deutsche Forschungsgemeinschaft (DFG, German Research Foundation) -- Project-ID 279384907 -- SFB 1245.
A.E. acknowledges the support from the DFG through grant LA 587/22-1.
E.O. is supported by the Swedish Research Council (Project No. 2020-00452).
The authors gratefully acknowledge the computing time provided to them on the high-performance computer Lichtenberg II at TU Darmstadt (Project-ID 2623), funded by the German Federal Ministry of Education and Research (BMBF) and the State of Hesse.
This research has made use of the Astrophysics Data System, funded by NASA under Cooperative Agreement 80NSSC25M7105.

Software: \flash{} \citep{Fryxell2000}, \nulib{} \citep{OConnor2015}, \winnet{} \citep{Reichert2023}, \code{Numpy} \citep{Harris2020}, \code{Matplotlib} \citep{Hunter2007}, \code{yt} \citep{Turk2011}, \code{h5py} \citep{Collette2013}

\section*{Data Availability}

The data related to this work will be shared upon reasonable requests to the authors.


\bibliographystyle{mnras}
\bibliography{refs}




\bsp	
\label{lastpage}
\end{document}